\documentclass[useAMS,graybox]{svmult}

\usepackage{mathptmx}       
\usepackage{amssymb}
\usepackage{helvet}         
\usepackage{courier}        
\usepackage{type1cm}        
\usepackage{makeidx}         
\usepackage{graphicx}        
\usepackage{multicol}        
\usepackage[bottom]{footmisc}

\makeindex             

\begin{document}

\title*{Clustered Star Formation: A Review}
\author{Richard J. Parker}
\institute{Richard J. Parker \at Institute for Astronomy, ETH Z{\"u}rich, Wolfgang-Pauli-Strasse 27, 8093, Z{\"u}rich, 
   Switzerland, \email{rparker@phys.ethz.ch}}
%
%
\maketitle


\abstract{In this contribution I present a review of star formation in clusters. I begin by discussing the various definitions of what constitutes a star cluster, and then compare the 
outcome of star formation (IMF, multiplicity, mass segregation and structure and morphology) in different star-forming regions. I also review recent numerical models of star formation in clusters, before ending with a summary of the potential effects of dynamical evolution in star clusters.}

\section{What is a star cluster?}
\label{intro}

A fundamental question regarding the nature of star formation is where do most stars form? The seminal review by Lada \& Lada (2003) suggested that 70 -- 90\% of stars form 
in embedded clusters, some of which remain bound and evolve to open clusters, and the remainder disperse and contribute to the Galactic field. In this picture, star clusters 
can be thought of as the fundamental `unit' of star formation. However, to describe clusters as a unit of star formation naturally requires a definition of what constitutes 
``clustered'' versus non--clustered, or ``isolated'' star formation. 

A recent study of YSOs in the local solar neighbourhood by Bressert et al (2010) showed that the distribution of stellar surface densities for 12 combined regions of star formation 
has a smooth, continuous distribution, with no distinction between isolated and clustered star formation. It is therefore problematic to define a cluster based on stellar 
surface density.

 A potential alternative is to apply a graph theory approach, in which all the stars in a field of view are uniquely linked by a minimum spanning tree (MST), which 
joins all the points via the shortest possible pathlength. One can then define a threshold length, above which branches in the MST a removed, leaving ``clusterings'' of stars 
(e.g. Gutermuth et al 2009; Schmeja 2011). 
Whilst this method is useful for picking out clusters against a crowded background, it requires the maximum threshold length to be defined (somewhat arbitrarily) and, as is 
often the case, the human eye is more efficient at picking out overdensities than the actual algorithms.

One can also define a star cluster based on the binding energy of a
group of stars. If the total binding energy is negative, then the
stellar system is bound and unlikely to fall 
apart through two-body relaxation. In an extension of this concept, Gieles \& Portegies Zwart (2010) made the distinction between a bound cluster and associations based on the 
crossing time of the star forming region (the time taken for a star to cross from one side to the other):
\begin{equation}
T_{\rm cr} = 7.5\left(\frac{R_{\rm eff}^3}{GM}\right)^{1/2},
\end{equation}  
where $R_{\rm eff}$ and $M$ are the effective half light radius, and mass of the star forming region, respectively. We obtain a value $\Pi$, by dividing the age of the region by 
$T_{\rm cr}$; regions with $\Pi > 1$ are bound clusters, and those with $\Pi < 1$ are unbound associations. 

As we have seen, actually defining what a cluster is can be difficult, and often is merely a matter of personal opinion. For the remainder of this review, I will focus more on 
asking the question of whether star formation is a universal process in different star-forming regions, and how we can frame this question. In order to do this, I will consider 
both diffuse and dense nearby star forming regions, which are ``clustered'' in the sense that they are an overdensity with respect to the Galactic field, but may not pass all, or 
any of the definitions above.

\section{Observations}
\label{observation}

The wealth of data from the recent Herschel observations (see contribution by J.\,Kirk) has shown star formation to be a highly filamentary process, and 
it is thought that star clusters may form at the intersection of filaments (``hubs'') where there are significant over-densities (Myers 2011). In the following subsections I will 
describe the observed outcome (i.e.\,\,the IMF, multiplicity, structure) of the star formation process in different regions (hereafter ``clusters'').

\subsection{The Initial Mass Function}

A great deal of effort has gone in to observing the Initial Mass Function (IMF) in various environments, principally to look for evidence of environmentally dependent variations. 
However, the IMF appears to be remarkably invariant; the same form is observed in dense clusters and sparse associations, and in open and globular clusters. All of these IMFs 
are also consistent with the mass function in the Galactic field (Bastian et al 2010). 

Whilst the IMF is largely invariant (certainly on Galactic scales), some star forming regions display peculiarities which may hint at subtle differences in the IMF. For example, the 
Taurus association contains an excess of K-type dwarfs compared to the more numerous M-dwarfs; and much effort is focusing on determining whether the substellar regime of 
the IMF is also invariant across different regions. 

Recent studies have also suggested that the IMF may vary as a function of distance from the cluster centre, with several regions (e.g. the ONC, IC348) displaying an excess of 
brown dwarfs at the periphery of the cluster compared to the centre. 

\subsection{Multiplicity}

If the IMF is invariant, then what may vary as a function of star formation event? A promising avenue to explore is stellar multiplicity, within which there are several parameters 
that can be accurately measured in clusters. In addition to the bare fraction of stars in multiple systems, one can compare the distributions of orbital parameters of binaries, 
such as the semi-major axis distribution and the companion mass ratio distribution. 

However, the process of comparing these distributions in different clusters is not straightforward. Firstly, comparable data are often drawn from different observational 
programmes; one has to be careful not to compare apples to oranges. Most nearby clusters have had their visual binaries observed; these are binaries with on-sky separations 
which are sufficiently large to resolve the primary and secondary stars into individual components, but not so large that the component stars cannot be distinguished from 
background cluster members. As the distance to nearby clusters varies, then one can observe binaries with closer and wider separations in nearby clusters. For example, in Taurus 
the observable separation range is 18 -- 1000\,au, whereas we are restricted to 62 -- 620\,au in the more distant ONC. To enable a fair comparison between the binaries in 
each region, we are forced to throw away from any analysis the extra systems in Taurus which lie outside the common 62 -- 620\,au range. Such surveys also are limited by a 
contrast range and are therefore sensitive to the flux ratios between the primary and secondary component of the binary.  

A recent synthesis of the available data on nearby clusters suggests that the binary fraction (for systems in the separation range 62 -- 620\,au) does not depend strongly on the 
density of that region (King et al 2012). Preliminary analysis also suggests that the separation distributions are not statistically distinguishable in this range. This is puzzling 
because the regions span several orders of magnitude in stellar density; even if only one region was dynamically active (the ONC), we would expect the separation distributions to 
have been altered to a different degree by interactions, the number of which is set by the density.

Comparison of the mass ratio distribution can also shed light on the binary formation process, and early work suggests that this distribution is flat in the Galactic field and in 
several clusters (see the contribution by M. Reggiani).

\subsection{Structure and morphology}

The first analysis of structure and morphology in star forming regions was conducted by Larson (1995), who looked at the two-point correlation function in Taurus. By plotting 
local surface density against distance to nearest neighbour, one can pick out the binary regime versus the general hierarchical structure in the cluster. Interestingly, the separation 
corresponding to the break in power law from the binary regime to the cluster regime corresponded to the Jeans length, suggesting some underlying physical process. The same  
break at the Jeans length was found in $\rho$~Oph and the Trapezium cluster, although the latter result was later disputed (Bate et al 1998). 

However, structure can be quantified in a more meaningful way using the $Q$-parameter (Cartwright \& Whitworth 2004), which divides the mean length of the MST of all stars in 
the cluster, $\bar{m}$ by the mean separation between stars in the cluster, $\bar{s}$;
\begin{equation}
Q = \frac{\bar{m}}{\bar{s}}.
\end{equation} 
When a cluster has substructure, $Q < 0.8$, whereas $Q > 0.8$ indicates a centrally concentrated cluster. Using the $Q$-parameter it is possible to infer either the fractal dimension of the structured cluster, or the density profile of the smooth cluster. Sanchez \& Alfaro (2009) have found that many young clusters display substructure, due to 
their low $Q$-parameters.

\subsection{Mass segregation}

Some young clusters (e.g. the ONC) are observed to be mass segregated, whereby the most massive stars reside at the cluster centre. Mass segregation can be quantified in 
several ways, such as an MST analysis (Allison et al 2009a; Olczak et al 2011), surface density as a function of mass (Maschberger \& Clarke 2011) or distance from the cluster 
centre compared to the median value (Kirk \& Myers 2010).

\section{Theoretical and numerical models}
\label{theory}

There are many different theories of star formation in clusters. In an ironic symmetry with the observations, many theoretical models are able to reproduce an IMF independent 
of a wide range of initial conditions. Here I mention two competing theories for massive star formation in clusters, for the reason that they predict very different (observable) outcomes for the mass of the remaining cluster of stars.

The monolithic collapse scenario (e.g. McKee \& Tan 2003; Krumholz et al 2005, 2009) predicts a top--down star formation from the collapse of a massive star forming core. The 
collapse and fragmentation of this core gives the IMF directly, which may vary depending on the exact details of the collapse. For example, if the core remains relatively warm, it 
may not fragment much, and could in principle lead to `isolated' O-star formation, where the O-star is surrounded by a few lower-mass stars. Recent observational studies by 
Lamb et al (2010) and Bressert et al (2012) suggest that some O-stars do indeed form in isolation. 

Alternatively, massive stars (and a surrounding cluster) can form via competitive accretion (Bonnell et al 2001, 2003; Smith et al 2009), where the cloud fragments into equal-mass seeds, which then accrete the remaining 
gas in varying proportions, so  as to produce the IMF. As the gas density is highest in the centre of the cluster, the central stars become the most massive and naturally explain the observed mass segregation in some clusters. This process suggests that the mass of the most massive star in a cluster would therefore depend on the cluster mass (due to the 
proportion of gas accreted by the massive stars). If this dependence is a fundamental outcome of star formation, then observationally we would expect that the most massive star should \emph{always} be proportional to the cluster mass, as claimed by Weidner et al (2010).  

Recent hydrodynamical simulations of star cluster formation by Bate (2009,2012) have also made concrete predictions for the primordial multiplicity of stars. From a collapsing 500\,M$_\odot$ 
cloud, 183 stars and brown dwarfs are formed (Bate 2012), with a mass function similar to the Chabrier system IMF. Additionally, the binary fraction is similar to that in the 
Galactic field, and decreases with decreasing primary mass. The distributions of the binary orbital parameters are also in remarkably good agreement with the field. 

\section{Dynamical evolution}
\label{dynamics}

Whilst the detailed hydrodynamical simulations of star formation make concrete predictions, they are extremely computationally expensive, and render a comparison of different 
(simulated) clusters impossible. For this reason, pure $N$-body simulations are also useful in assessing the change in cluster morphology and structure, and the effects 
of the cluster environment on the outcome of star and planet formation (e.g. primordial multiplicity and protoplanetary discs).

The first simulations to look at the effect of cluster dynamics on primordial binaries were pioneered by Kroupa (1995) and Kroupa et al (1999) who used an ``inverse population 
synthesis'' to infer the dynamical processes which affect binaries as their host clusters evaporate into the field. More recent simulations have looked at the evolution of 
cluster structure and morphology. Simulations of subvirial (cool) clusters with structure have shown that dynamical mass segregation and the formation of Trapezium-like 
systems can occur on short timescales (Allison et al 2009b, 2010; Allison \& Goodwin 2011). The evolution of substructure in clusters is particularly interesting as in addition to 
cool clusters, both virial (`tepid') and supervirial (`warm') clusters can process primordial binaries and planetary systems (Parker et al 2011; Parker \& Quanz 2012), even if they 
do not attain high global densities.\\

To briefly summarise, star clusters can tell us much about star formation and in addition to IMFs, we should be comparing multiplicity properties, and the internal and global structure of clusters to search for fundamental differences in the star formation process as a function of environment.


\begin{acknowledgement}
I wish to thank the SOC for inviting me to present this review, and the LOC for organising such an enjoyable conference in a beautiful location. I would also like to thank Ant~Whitworth for many entertaining and enlightening conversations (usually over beer!).
\end{acknowledgement}
%

%
%
%
\biblstarthook{}



\end{document}